\documentclass[prd,nofootinbib,preprint,floatfix]{revtex4}

\usepackage{amsmath,amssymb}
\usepackage[normalem]{ulem}
\usepackage{graphicx}
\usepackage{array}
\usepackage{color}

\begin{document}

\vspace*{.25in}
\title{Possible scenario for MaVaN's as the only neutrino 
flavor conversion mechanism in the Sun}

\author{P.~C.~de Holanda} 
\email{holanda@ifi.unicamp.br}
\affiliation{
Instituto de F\'{\i}sica ``Gleb Wataghin'' \\ 
Universidade Estadual de Campinas, UNICAMP \\
13083-970, Campinas, S\~ao Paulo, Brasil
}

\begin{abstract}
Mass Varying neutrino mechanisms were proposed to link the neutrino 
mass scale with dark energy, addressing the coincidence problem. In some 
scenarios this mass can present a dependence on the baryonic 
density felt by neutrinos, creating an effective neutrino mass that depends 
both on the neutrino and baryonic densities. 
In this article we investigate the possibility that a neutrino effective 
mass is the only flavour conversion mechanism acting in neutrino oscillation 
experiments. We present a parameterization on the environmental effects on 
neutrino mass that produces the right flavour conversion probabilities for 
solar and terrestrial neutrinos experiments.

\end{abstract}
\maketitle

\section{Introduction}

The proposal that Mass Varying Neutrinos are coupled to dark 
energy~\cite{dark1} results in a fluid with negative pressure that could 
mimic the effects of a cosmological constant and induce a cosmic acceleration. 
The cosmological consequences of such coupling were widely 
addressed in a number of papers
~\cite{Hung:2003jb,Peccei:2004sz,Li:2004tq,Bi:2004ns,Brookfield:2005td,Horvat:2005ua,Afshordi:2005ym,Li:2005zd,Weiner:2005ac,Brookfield:2005bz,Takahashi:2006jt,Ringwald:2006ks,Zhao:2006zf,Bjaelde:2007ki,Ichiki:2007ng,Ichiki:2008rh,Bjaelde:2008yd}.

Effects on neutrino 
oscillation~\cite{Kaplan:2004dq,Schwetz:2005fy,Gu:2005pq,Gu:2005eq,Barger:2005mn,Barger:2005mh,Zurek,GonzalezGarcia:2005xu,Cirelli:2005sg,Abe:2008zza}
were analyzed in different scenarios. 
In particular, it was proposed that the Mass Varying Neutrinos phenomenology 
could lead to 
a neutrino mass dependence with baryonic density~\cite{Kaplan:2004dq} through 
non-renormalizable operators that couple the acceleron to the
baryonic matter.
In ref.~\cite{GonzalezGarcia:2005xu} some limits on the product of
the effective neutrino-scalar and matter-scalar Yukawa couplings were obtained 
by comparing the solar neutrino and KamLAND data, assuming that 
MaVaN's  mechanism plays a sub-leading role in neutrino flavour 
conversion.

We present in this article a possible scenario where the neutrinos have a
vanishing mass and mixing in vacuum. The positive oscillation
indications in terrestrial and solar neutrino data are fully explained due
to environmental effects that generate an effective mass and mixing in
presence of baryonic matter.  We perform an analysis
of solar 
data~\cite{chlorine,sage,Altmann:2005ix,gallex,sk,sno,Aharmim:2005gt,Aharmim:2008kc,Arpesella:2008mt}, 
for one specific choice of parameters, 
obtaining an acceptabe solution to solar neutrino problem. 

In Sec.~\ref{sec:mechanism}, we introduce the general theoretical 
framework which we will consider in this paper. 
In Sec.~\ref{sec:terrestrial} we show how reactor and accelerator neutrino 
experiments test only the effective neutrino mass.
In Sec.~\ref{sec:solar} we discuss the neutrino flavour conversion probability 
in the Sun for this mechanism. 
In Sec.~\ref{sec:atmospheric} we address the problem for atmospheric neutrinos.
Finally, in Sec.~\ref{sec:conclusion}, we discuss our results and summarize 
our conclusions.

\section{MaVaN's mechanism and parameterization}
\label{sec:mechanism}

In a previous work~\cite{GonzalezGarcia:2005xu} 
we found limits for the product of
the effective neutrino-scalar and matter-scalar Yukawa couplings
described in~\cite{Kaplan:2004dq}. But an assumption about the
adiabaticity of the transition was made, and with this assumption
we found that the new physics evoked always plays the role of a
sub-leading effect compared to the standard oscillation scenario. 

The aim of this work is try to find if there is at least one
combination of parameters for new physics that could lead to an
acceptable solution to the neutrino oscillation data where such new physics is 
more then a sub-leading effect. This implies
that non-adiabatic effects would be present in solar neutrino
evolution. 

A particular case from this oscillation plus MaVaN's scenario would be the
extreme opposite from the one investigated previously, {\it i.e.} the
situation where MaVaN's physics is the main flavour conversion mechanism in
terrestrial experiments. This is the scenario we investigate in this paper, 
where the new physics generates the neutrino mass
for all terrestrial indications for neutrino oscillations, and
is present in all three neutrino families.

The mass matrix in flavour basis has the form:
\begin{equation}
M= 
U_M
\left(
\begin{array}{ccc}
M_{1} &  & \\
 & M_{2} &  \\
 &  & M_{3} \\
\end{array}
\right)
U_M^{-1}
+
U_v
\left(
\begin{array}{ccc}
m_{1} &  & \\
 & m_{2} &  \\
 &  & m_{3} \\
\end{array}
\right)
U_v^{-1}
\label{eq:massmatrix3fam}
\end{equation}
where $U_M$ is the mixing matrix and  $M_i$ are the mass eigenvalues
related to MaVaN's effect, while $U_v$ and $m_i$ are the mixing matrix 
and mass eigenvalues in vacuum. We assume that the mixing angles 
due to MaVaN's effects are constants.

The environment effect is introduced as a dependence of
the mass terms with the baryonic matter density with the following 
parameterization:
\begin{equation}
M_{i}=M_0 \tanh{\left(\lambda_i
  \frac{\rho}{3\,{\rm g/cm^3}}\right)} 
\label{eq:massterm}
\end{equation}
This parameterization was chosen to reproduce two features that we want
our mass matrix to present:
\begin{enumerate}
\item a linear growth of mass with baryonic density for small values
  of this density. This is the behavior suggested in~\cite{Kaplan:2004dq}, 
assuming 
  a small shift in the value of $A$ with respect to its ground value.
\item a saturation of the environmental dependence of neutrino masses
  for large values of the baryonic density.
\end{enumerate}

\section{Reactor and accelerator neutrinos}
\label{sec:terrestrial}

If we assume a constant Earth density in the crust, the neutrino oscillation
probabilities can be written analytically, with the same form of the 
the vacuum oscillation
probabilities in the known mass-induced oscillation scenario. 
The standard mass and mixing angles are replaced by the effective mass and
mixing angles in matter.

The positive indications for neutrino oscillation in Earth
experiments~\cite{k2k,minos,kamland,chooz} tell us that the mixing
angles that diagonalize this mass matrix in presence of Earth matter are:
\begin{eqnarray}
&& \sin^22\theta_{23}>0.90~~(90\%~{\rm C.L.})  \nonumber \\
&& 0.4 < \tan^2\tilde{\theta}_{12} < 2.4 ~~(95\%~{\rm C.L.})\nonumber \\
&& \sin^2\tilde{2\theta}_{13} < 0.1 ~~(90\%~{\rm C.L.})
\end{eqnarray}
with the following mass squared differences:
\begin{equation}
  \Delta \tilde{m}^2_{21}\approx 7.6\times 10^{-5} {\rm eV}^2 ~~~;~~~
\Delta \tilde{m}^2_{31}\approx 2.5\times 10^{-3} {\rm eV}^2
\label{eq:deltam}
\end{equation}

We want to investigate the scenario where the MaVaN's are the leading
neutrino flavour conversion mechanism, so we assume that the
vacuum mass eigenvalues are very small, $\Delta m^2_{ij}<<10^{-5}$ eV$^2$, 
and all the flavour conversion is induced by the first term in right-hand 
side of Eq.~\ref{eq:massmatrix3fam}.

Assuming a constant Earth crust density of $\rho \sim 3$ g/cm$^3$, we 
tune our parameters in eq.~\ref{eq:massterm} in order to reproduce the 
above $\Delta \tilde{m}^2$. 
This can be achieved for instance with the following 
choice:
\begin{eqnarray}
\lambda_1=0~~~;~~~&\lambda_2=0.18 &~~~;~~~ \lambda_3=10 \nonumber \\
&M_0=5\times10^{-2} {\rm eV}& 
\end{eqnarray}

With this parameterization the mass of the
second family
is close to a linear regime for the baryonic densities 
present at Earth, and for the crust density of $\rho\sim 3$ g/cm$^3$ we 
obtain $\Delta\tilde{m}^2_{21}=7.9\times 10^{5}$ eV$^2$, with a 
$\rho^2$ dependence. 
For the atmospheric neutrino scale, the third mass eigenvalue is already 
saturated for the Earth crust density, leading to 
$\Delta\tilde{m}^2_{32}=2.4\times 10^{3}$ eV$^2$.

Choosing a convenient set of mixing angles, 
the oscillation of terrestrial neutrino experiments are satisfactorily 
explained by this parameterization.

\section{Solar neutrinos}
\label{sec:solar}

For solar neutrinos, besides the environment
effect in the mass matrix, the standard matter interaction term 
will have an important role in neutrino conversion. 
The evolution matrix has the form:
\begin{equation} 
i \frac{d}{dr} 
\left(\begin{array}{c} \nu_e \\
\nu_\mu \\ \nu_\tau    \end{array}\right)=
\left[
\frac{1}{2E_\nu}MM^t +
\left(    \begin{array}{ccc}
        V_{\rm CC}(r) & ~0 & ~0\\
        \hphantom{-} 0 & ~0 & ~0 \\
        \hphantom{-} 0 & ~0 & ~0 \\
    \end{array}\right)\right]
\left(\begin{array}{c} \nu_e \\
\nu_\mu  \\ \nu_\tau   \end{array}\right),
\label{eq:evol}
\end{equation}
where $M$ is the mass matrix of Eq.~\ref{eq:massmatrix3fam}.

The standard interaction term would be negligible
at solar neutrino production point for the energies of interest, 
indicating that 
the mixing matrix that diagonalizes the full evolution matrix
would be the same in KamLAND and in the center of the Sun.

However, the mass term decreases faster than 
the standard interaction one as the neutrino travels towards the Sun surface, 
and these terms become of the same order 
around $r_{Sun}\sim 0.8$ for typical solar neutrino energies. The 
mixing matrix would start
to feel the modifications due to the standard interaction term $V_{CC}$
and the vacuum mass terms as the neutrino approaches the surface of the Sun.

\begin{figure}
\includegraphics[width=12cm]{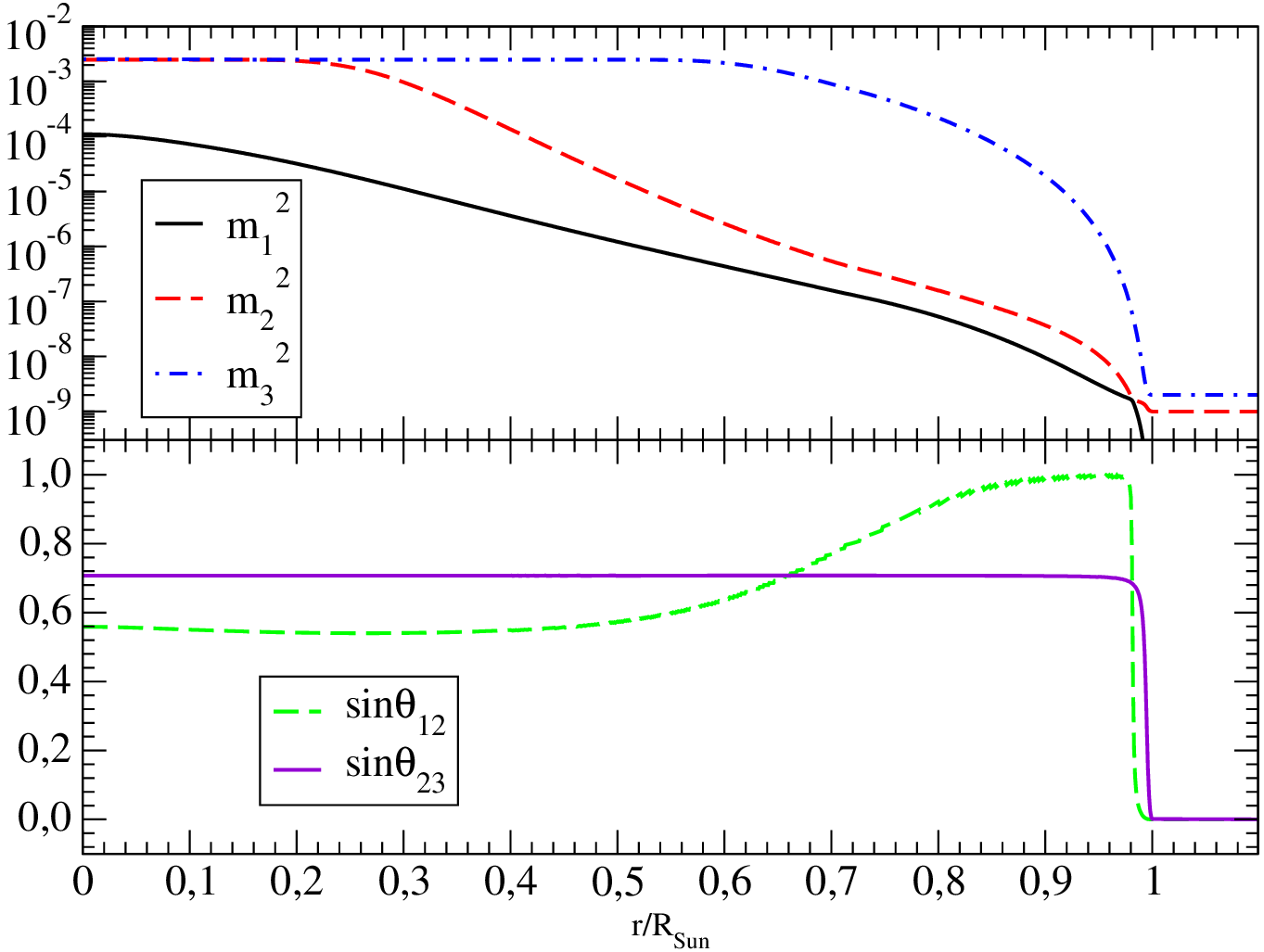}
\caption{Mass eigenvalues and mixing angles for $E_\nu=10$ MeV.}
\label{fig:mass1}
\end{figure}

\begin{figure}
\includegraphics[width=12cm]{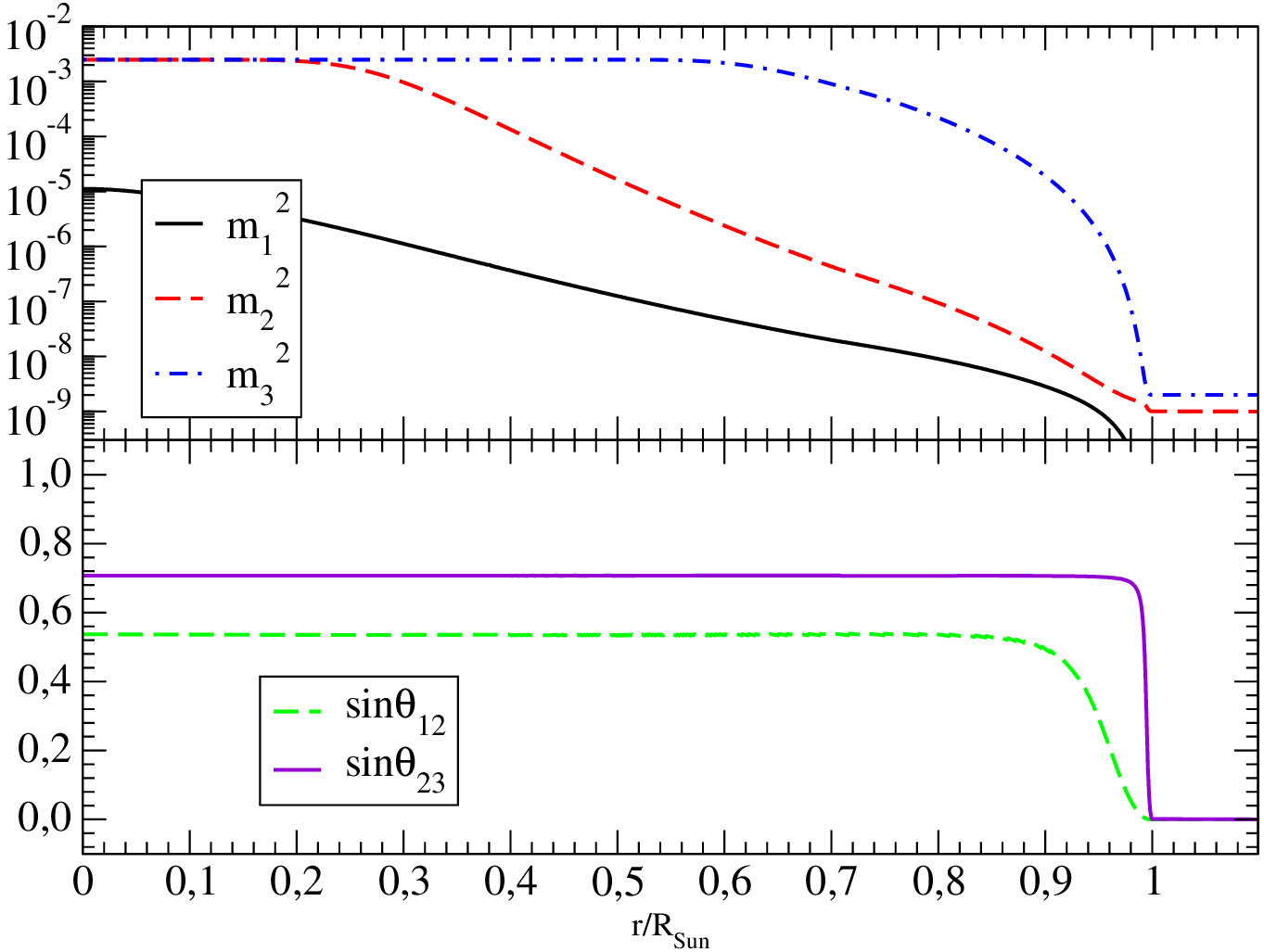}
\caption{Mass eigenvalues and mixing angles for $E_\nu=1$ MeV}
\label{fig:mass2}
\end{figure}

In Fig.~\ref{fig:mass1} and Fig.~\ref{fig:mass2} we can see the behavior 
of the eigenvalues of the evolution matrix and the mixing angles inside 
the sun, for energies of 10 MeV and 1 MeV, respectively. We have chosen for 
the vacuum parameters:
\begin{eqnarray}
&&\theta_{13}=\theta_{23}=\theta_{12}=0 \nonumber\\
&&\Delta m^2_{21}=10^{-9}~{\rm eV}^2~~~;~~~
\Delta m^2_{32}=2\times 10^{-9}~{\rm eV}^2
\end{eqnarray}

In this scenario the third
family would assume a constant value for its mass eigenstate around 
$r\sim 0.6$, while the second family 
would achieve the saturation value around $r \sim 0.2$, corresponding 
to baryonic densities of $0.4$ and $30$ g/cm$^3$, respectively.

We can estimate the survival neutrino probability assuming that the 
transition is adiabatic until very close to the solar surface, and then 
the neutrino assumes the vacuum values of mass and mixing in an extremely 
non-adiabatic transition. The electron neutrino is created as an admixture 
of the two first mass eigenstates, 
\[
|\nu_e>=\cos\theta_{KL}|\nu_1>+\sin\theta_{KL}|\nu_2>
\]
where I wrote $\theta_{KL}$ to make it explicit that 
the mixing angle measured in KamLAND is the same 
at Sun's interior, $\theta_{KL}=\tilde{\theta}_{12}(r=0)$. For $E=10$ MeV 
(fig~\ref{fig:mass1}) the mixing angle increases as the neutrino travels 
towards the Sun's surface due to $V_{CC}$, and at 
$r\sim 0.95$ the mixing angle achieves the value $\sin\tilde{\theta}_{12}=1$. 
Assuming this change to be adiabatic, the probability to have an 
electron neutrino at this point can be estimated to be:
\[
P_{ee}=\sin^2\theta_{KL}\sin^2\tilde{\theta}_{12}+
\cos^2\theta_{KL}\cos^2\tilde{\theta}_{12}=\sin^2\theta_{KL}\sim 0.3~~~,
\]
and the conversion probabilities are:
\[
P_{e\mu}=P_{e\tau}\sim 0.35
\]
Since in an extremely non-adiabatic transition there is no conversion between 
flavour eigenstates, these are the conversion probabilities right outside 
the Sun. Assuming no mixing in vacuum,
these probabilities correspond to the probabilities measured at Earth, and 
to the admixture of mass eigenstates that evolve from Sun to Earth.

The Earth regeneration is expected to be small in this scenario.
The key feature is that the vacuum mixing angle $\theta_{23}$ was
made very small, while $\tilde{\theta}_{23}$ is maximum close to 
solar border. So the extremely non-adiabatic transition from the outer 
parts of the Sun to the vacuum leads to a strong production of the mass
eigenstate $\nu_3$. And since we have
$P_{ee}\sim P_{e\mu}\sim P_{e\tau}\sim 1/3$  the non-adiabatic
transition to vacuum will lead to an equipartition of $\nu_1$,
$\nu_2$ and $\nu_3$, and then no regeneration effect at Earth. So the
absence of Earth regeneration in Super-Kamiokande and SNO is a direct
consequence of the probability $P_{ee}\sim 1/3$ inside the Sun.

Doing the same analysis for $E=1$ MeV, we can estimate the probabilities 
the same way, with the difference that the mixing angle just before the 
non-adiabatic transition to vacuum is not changed by standard interaction 
term. Then: 
\[
P_{ee}=\sin^4\theta_{KL}+\cos^4\theta_{KL}\sim 0.6
\]
and
\[
P_{e\mu}=P_{e\tau}\sim 0.2
\]

Again these probabilities correspond to the probabilities at Earth if we 
assume non-adiabatic transitions and no vacuum mixing. 
For these energies we would expect a strong 
effect due to interaction with Earth matter, but since $P_{\nu_1}>P_{\nu_2}$, 
the Earth effect would lead to a stronger conversion at night.

We numerically calculated the solar neutrino survival probability for 
the values of $\tan^2\tilde{\theta}_{12}=0.4$, $\tan\tilde{\theta}_{23}=1$ and
$\tilde{\theta}_{13}=0$ for the mixing angles. 
For the vacuum parameters, we assumed
very small values for the mass eigenvalues and mixing angles.

\begin{figure}
\includegraphics[width=12cm]{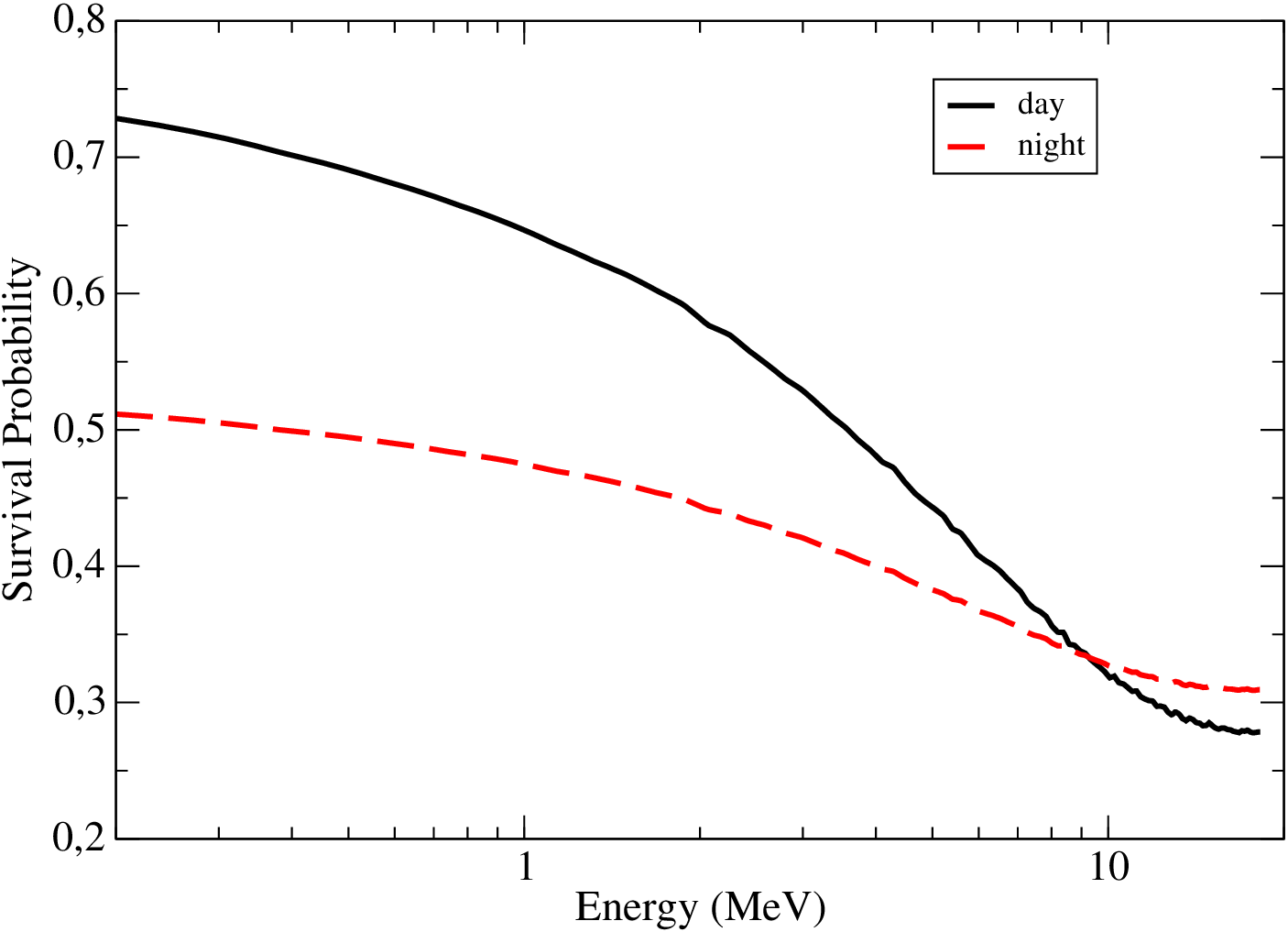}
\caption{Solar neutrinos survival probability during the day (solid
  line) and the night (dashed line).}
\label{fig:prob}
\end{figure}

The result of our calculation can be seen
in Fig.~\ref{fig:prob}. All energy dependence present in the
probability comes from non-adiabatic effects in neutrino evolution
close to the border of the Sun.

The probability obtained reproduces two main ingredients of the
desired conversion to explain the solar neutrino data: a higher
survival probability for low energy neutrinos and a small regeneration
effect for high energy neutrinos. Besides, there is one interesting
feature of this mechanism that is the positive day-night asymmetry for
low energy solar neutrinos, which could be tested by Borexino. Also, 
this would lead to a negative
winter-summer asymmetry in low energy solar neutrino experiments due to
the difference of day and night duration in winter and summer. GNO
recently reported a winter-summer asymmetry of $\Delta(W-S)=-7.6\pm
8.4$ SNU ($\sim -11\%$) in their full data analysis. 
The expected value from the $1/d^2$ modulation only is $+2.3$ SNU 
($+3.3\%$), slightly  $1\sigma$ above GNO results.

\section{Numerical Analysis}
\label{sec:numerical}

We present in this section the results of the numerical analysis of
solar data for the parameterization discussed in the previous 
section.
Details of our solar neutrino analysis have been described in previous
papers~\cite{oursolar,pedrosolar,GonzalezGarcia:2005xu}. 
We use the solar fluxes from Bahcall and Serenelli (2005)~\cite{BS05}. 
In comparison with previous works, we include the new SNO 
data~\cite{Aharmim:2008kc}, and included latest Borexino 
results~\cite{Arpesella:2008mt}. Besides, 
Gallex/GNO results were split by winter and summer data. 
The solar neutrino data includes a total of 124 data points: 

\begin{itemize}
\item 1 data point for Homestake results~\cite{chlorine}.
\item 1 data point for SAGE results~\cite{sage}.
\item 2 data points for Gallex/GNO results, for winter and 
summer~\cite{gallex,Altmann:2005ix}.
\item 44 data points for Super-Kamiokande zenithal/spectral 
bins~\cite{sk}.
\item 34 data points for SNO, phase 1~\cite{sno}.
\item 38 points for SNO, phase 2~\cite{Aharmim:2005gt}.
\item 1 point for Borexino data~\cite{Arpesella:2008mt}
\item 3 points for SNO, phase 3~\cite{Aharmim:2008kc}
\end{itemize}


We obtaine a viable solution to solar neutrino problem with the
following parameters.
\[
\tan^2\theta_{KL}=0.4 ~~~~ 
\Delta m^2_{KL} = 7.9 \times 10^{-5}\,{\rm eV}^2~~,
\]
with a $\chi^2=118.9$.
For comparison, our standard analysis provides a $\chi^2=113$ for the 
same parameters values. 

As mentioned before, a clear signature of this mechanism would be the
day-night asymmetry for low energy neutrinos. For the point specified
above, the Berilium line neutrinos would have:
\[
P_{day}=0.62 ~~~ P_{night}=0.45\,,
\]
leading to a day-night asymmetry in Borexino of $A_{DN}=30\%$.

For the winter-summer asymmetry in low-energy experiments,
this same point would predict:
\[
R_{summer}=68.0 ~~~ R_{winter}=68.7 \,,
\]
leading to a winter-summer asymmetry of $A_{WS}=+1\% $.
MSW prediction for such asymmetry at the b.f.p. is around $A_{WS}=+4\%$.

\section{Atmospheric neutrinos}
\label{sec:atmospheric}

In this scenario, atmospheric neutrinos would oscillate through an
almost constant $\Delta m^2$ inside the Earth, with a small decrease
in its value in Earth's core due to the rise of the second mass
scale with higher densities. The possibility of environmental effects on 
atmospheric neutrinos were analyzed~\cite{Abe:2008zza} 
with other parameterization of 
the matter-dependence. In that work the following choice:
\[
\Delta m^2_{eff}=(1.95\times 10^{-3}) 
\left(\frac{\rho_e}{\rho_0}\right)^{-0.04}~{\rm eV}^2
\]
led to an acceptable solution to the SK atmospheric neutrino data. Here 
$\rho_e$ is the electron neutrino density and $\rho_0=6.02\times 10^{23}$ 
cm$^{-3}$. 

\begin{figure}
\includegraphics[width=12cm]{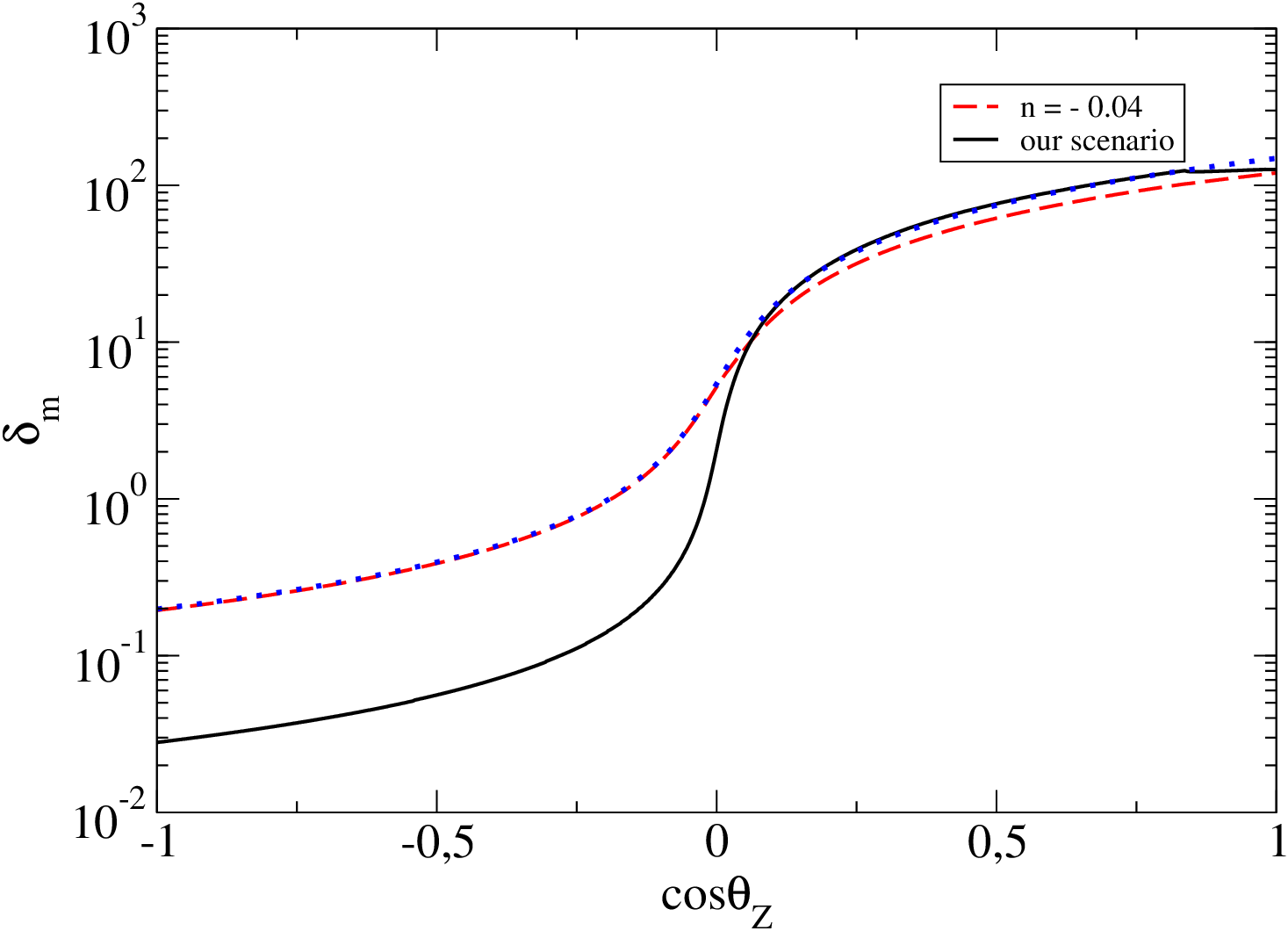}
\caption{Number of oscillation lengths for atmospheric neutrinos for our 
scenario (continuous line), standard oscillation (dotted line) and 
previous parametrization on MaVaN's mechanism (dashed line).}
\label{fig:mass.atm}
\end{figure}

To compare our proposal with the one above, we plot in 
fig.~\ref{fig:mass.atm} the number of oscillation lengths covered by a 
typical atmospheric neutrino, given by:
\[
\delta_m=\frac{\int \Delta m^2_{eff} dL}{1~{\rm GeV}}~~~.
\]
The dashed line represents the choice in~\cite{Abe:2008zza}, and the solid 
line the $\delta_m$ obtained with our parameterization. Also presented in 
dotted line is the 
standard oscillation scenario with constant $\Delta m^2=2.3\times 10^{-3}$ 
eV$^2$.

We can see that the main difference between our choice of parameterization 
and the standard scenario happens for down-going neutrinos, with 
intermediate energy, $E_\nu\sim {\rm few}$ GeV. There are some indications 
that these neutrinos are already oscillating, which would decrease the 
concordance of this model with atmospheric neutrino oscillation data.
But due to the very similar agreement with standard oscillation in 
a large range of atmospheric neutrino flux parameters, we believe that an 
overall fit would give a viable solution also to atmospheric neutrinos. 
A detailed numerical analysis would be necessary to verify this issue.

\section{Conclusions}
\label{sec:conclusion}

We present in this article a possibility where all flavour conversion on 
neutrinos can come from environmental effects. The standard oscillation 
scenario is still the most elegant theoretical framework that explains 
neutrino flavour conversion, not only due to the excellent numerical fit to 
all oscillation data, but also due to the success of the research program 
that predicted, for instance, the correct signal at KamLAND based on one 
possible solution to the solar neutrino problem. The model proposed here 
explains such concordance {\it a posteriori}. However, it also 
make some very particular predictions on future experiments, that can be 
verified in the close future. Borexino experiment is already taking data, 
and very soon can test a possible day-night asymmetry for low energy solar 
neutrinos. The author does not know any other model that predicts such 
asymmetry, making this prediction an unique signature of MaVaN's as 
the mechanism on neutrino flavour conversion.

\begin{acknowledgments}
The author is thankful to R. Zukanovich-Funchal, M. C. Gonzalez-Garcia and 
M. Maltoni for fruitful discussions, and for CNPq and Fapesp for 
financial support.
\end{acknowledgments}

\end{document}